# Variation of the fundamental band gap nature in curved two-dimensional WS$_2$

E. Blundo[‡], M. Felici[‡], T. Yildirim[†], G. Pettinari[◊], D. Tedeschi[‡], A. Miriametro[‡], B. Liu[†], W. Ma[†], Y. Lu,[†#], and A. Polimeni[‡,*]

[‡] Dipartimento di Fisica, Sapienza Università di Roma, 00185 Roma, Italy

[†] Research School of Electrical, Energy and Materials Engineering, College of Engineering and Computer Science, The Australian National University, Canberra, ACT 2601, Australia

[◊] Institute for Photonics and Nanotechnologies, National Research Council, 00156 Roma, Italy

[#] ARC Centre of Excellence in Future Low-Energy Electronics Technologies (FLEET) ANU node, Canberra, ACT 2601, Australia

We report a strain-induced direct-to-indirect band gap transition in mechanically deformed WS$_2$ monolayers (MLs). The necessary amount of strain is attained by proton irradiation of bulk WS$_2$ and the ensuing formation of one-ML-thick, H$_2$-filled domes. The electronic properties of the curved MLs are mapped by spatially- and time-resolved micro-photoluminescence revealing the mechanical stress conditions that trigger the variation of the band gap character. This general phenomenon, also observed in MoS$_2$ and WSe$_2$, further increases our understanding of the electronic structure of transition metal dichalcogenide MLs and holds a great relevance for their optoelectronic applications.

The properties of solids are very sensitive to variations in bond length ensuing a mechanical deformation or stress. This is especially true in the case of two-dimensional (2D) crystals (such as graphene, hexagonal-BN and transition metal dichalcogenide, TMD, monolayers, MLs) due to their all-surface nature[1]. Particularly appealing in TMD MLs is the strong coupling between the valley/spin/orbital degrees of freedom and the lattice structure, reflected in the strong response of their electronic[2], transport[3], and optical[4] properties to strain. In particular, non-uniform strains turn out to be extremely relevant: On the one hand, strain gradients in 2D-TMDs can lead to a coherent drift of photo-generated carriers, relevant for photon harvesting[5,6]. On the other hand, a non-uniform strain gives rise to pseudo-electromagnetic fields enabling the observation of novel transport phenomena[7].

In this letter, we report a study of the band gap character in mechanically-deformed $WS_2$ 2D crystals. The deformation follows the local blistering over a micron-sized region of the upper layer of bulk $WS_2$ flakes exposed to proton irradiation[8]. The resulting spherically-shaped MLs, hereafter named domes, host non-uniform and high strain fields, evaluated by finite-element method (FEM) calculations and consistently compared with micro-Raman measurements. Steady-state and time-resolved micro-photoluminescence (PL) mapping of the band gap states over the surface of a single dome unveils dramatic changes in the emission energy, intensity and decay time. Such changes are related to the built-in tensile strain of the dome and are ascribed to a strain-induced direct ($K_{CB}$-$K_{VB}$)-to-indirect ($K_{CB}$-$\Gamma_{VB}$) band gap transition (CB and VB stand for conduction and valence band, respectively). The strain conditions that determine the cross-over of the VB $\Gamma$ and K states are found. A similar behavior is also observed in $MoS_2$ and $WSe_2$.

The dome-shaped membranes under study were created from bulk $MX_2$ flakes (where M: W or Mo, and X: S, Se or Te), which were mechanically exfoliated on Si substrates and then proton-irradiated with a low-energy beam[8,9]. Here, we focus on $WS_2$. The accelerated protons penetrate through the flake surface and $H_2$ forms just one layer beneath the surface, as described in Ref. 8. As a consequence of the balance between the gas expansion, the van der Waals forces holding the S-W-S planes together and the material's elastic properties, localized swelling of just one ML takes place, resulting in the formation of atomically-thin and spherically-shaped domes [see the atomic force microscope, AFM, image in Fig. 1(a)]. The domes cage highly-pressurized $H_2$ and are durable thanks to the impermeability to $H_2$ of TMD MLs[10]. As shown in Fig. 1(a), domes with different size stud the flake's surface, nevertheless featuring an aspect ratio, $h_m/R$=0.16±0.02, independent of $R$[8,11] ($R$ is the footprint radius and $h_m$ is the maximum height of the domes). The domes were characterized by micro-Raman and micro-PL experiments using a 532 nm laser as the excitation source. A diffraction grating monochromator coupled to a Si-CCD was used for spectral analysis of the signal. Time-resolved micro-PL was performed using a supercontinuum laser tuned at 532 nm with ~50 ps pulse width and 77.8 MHz repetition rate. The signal was time-analyzed by a Si APD with 250 ps temporal resolution. Spatially-resolved optical measurements were performed in back-scattering configuration via a 100× objective with NA=0.9, resulting in a laser spot with standard deviation equal to 0.23±0.01 μm[12]. Finally, the strain tensor over the dome's surface was computed via FEM calculations.

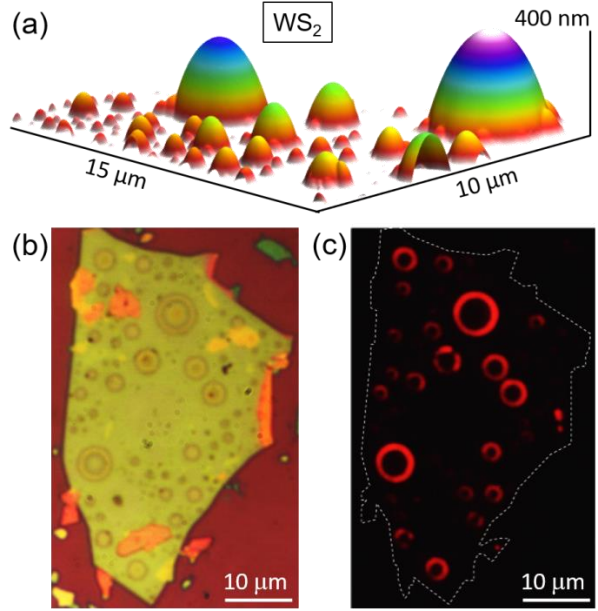

FIG. 1. (a) 3D AFM image of a bulk $WS_2$ flake irradiated with protons (dose $d_H=4\times10^{16}$ protons/cm$^2$), showing the formation of almost-perfectly-spherical domes. (b) Optical image of a $WS_2$ flake, where many relatively large domes formed after proton-irradiation ($d_H=5\times10^{16}$ protons/cm$^2$). (c) Laser-excited red luminescence coming from the same flake shown in panel (b).

Fig. 1(b) shows the optical microscope image of a proton-irradiated $WS_2$ flake acquired with a 50× (NA=0.5) objective. The outer circular borders locate the domes' footprint while the internal patterns of each dome are due to interference effects between the light reflected by the dome top surface and the flat $WS_2$ flake underneath[8]. Fig. 1(c) shows an optical image of the same flake, excited by a defocused 532 nm laser. The image was acquired at room temperature by filtering the laser out, thus letting the red luminescence (at ~690 nm) generated by the domes be revealed. Peculiarly, the brightly emitting region is restricted to an outer ring-like area independently of the dome footprint. This excludes interference (which would be largely dependent on the dome size) to be at the origin of the observed luminescence pattern, differently from what recently reported in $WS_2$ bubbles obtained after annealing of chemical-vapor-deposition-grown MLs[13]. In that work, interference effects are likely enhanced by the $SiO_2$ substrate located right beneath the ML bubbles and strongly modulate the emission. On the contrary, in the present case, the peculiar ring-like emitting area stems from the strain field acting over the domes, as detailed in the following.

To model the spatial evolution of the strain tensor and the height profile of the domes we performed FEM calculations within the framework of the nonlinear membrane theory[14,15,16]. The AFM-derived radius and height of the domes and the elastic properties of the material were used as input parameters. Fig. 2(a) (left axis) successfully compares the experimental (circles) and calculated (solid line) height profile along a radius ($0 \leq r \leq R$, where $r$ is the position with respect to the center) of the $WS_2$ dome, whose AFM image is shown in the inset. The right axis of Fig. 2(a) displays the calculated $r$ dependence of the principal components of the strain tensor — namely, along the circumferential ($\varepsilon_t$) and radial ($\varepsilon_r$) in-plane directions[14] and along the perpendicular ($\varepsilon_z$) out-of-plane direction. At the dome's summit, the (tensile) strain is isotropic biaxial ($\varepsilon_t = \varepsilon_r = 2.09\%$) in agreement with Hencky's model[4,14,16,17], whereas at the dome edges —where $\varepsilon_t = 0$— strain is uniaxial. The negative value of $\varepsilon_z$ all over the surface is caused by the membrane thinning following the in-plane tensile strain. The strain field across the dome is expected to induce remarkable changes in the electronic properties of the curved TMD membrane[5,18,19,20,21,22,23,24,25], giving rise to the peculiar phenomenology displayed in Fig. 1(c). Spatially-resolved micro-PL/Raman measurements were then performed on the dome shown in Fig. 2(a), which was chosen since its size ($R=2.85$ µm) is much larger than the probing laser spot (0.23 µm)[12], thus minimizing diffraction effects. However, it is important to note that the dome aspect ratio and, consequently, the strain distribution remain unchanged with the dome size[8] ensuring the general significance of the following results.

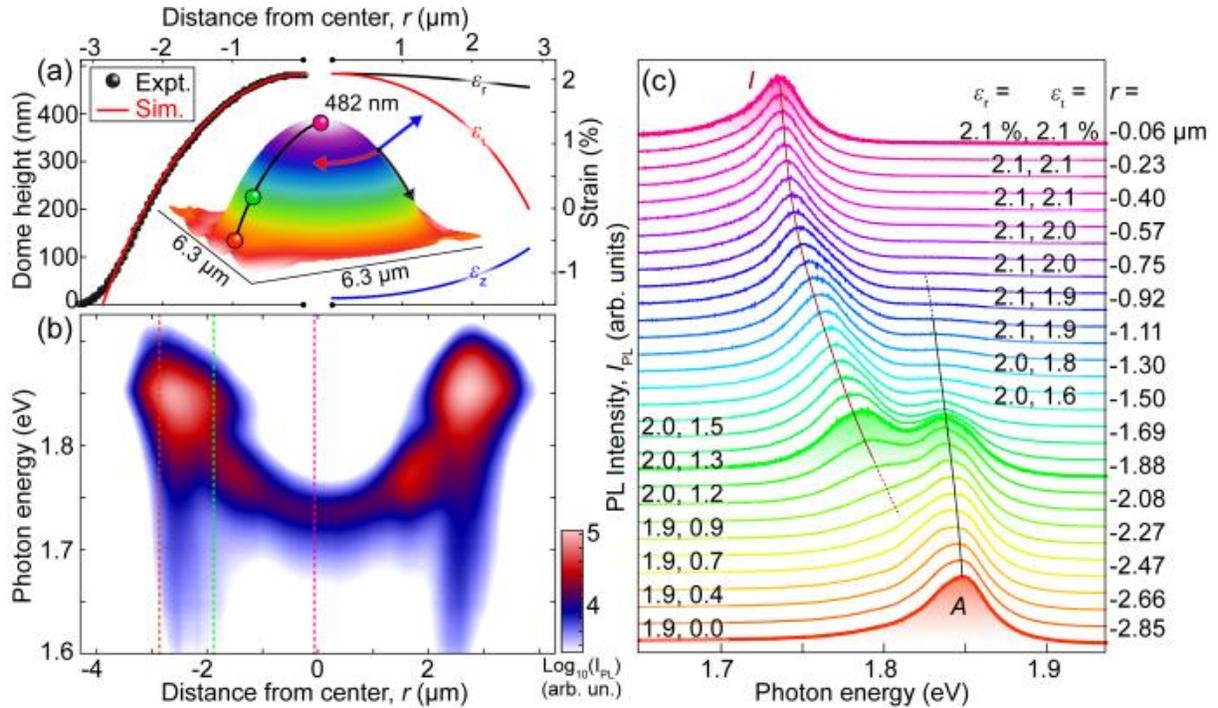

FIG.2. (a) Left: Height profile of a WS$_2$ dome formed after irradiation with $4\times10^{16}$ protons/cm$^2$, measured by AFM (black dots; the AFM image is shown as inset), and computed by FEM calculations (solid red line). Right: Dependence along the dome radius of the strain tensor components, represented as color-coded arrows in the inset. The three dots (purple: top; green: intermediate; orange: edge) correspond to the positions displayed as dashed lines in (b) and to the shadowed spectra in (c). (b) Micro-PL scan along a diameter of the dome displayed in (a), performed at 297 K. The horizontal axis indicates the laser spot position with respect to the dome center, the vertical axis indicates the emitted photon energy. The base-10 logarithm of the micro-PL intensity is shown in a false color scale. (c) Normalized emission spectra of the dome as the laser spot is scanned from the dome's left edge (bottom) to its apex (top). Spectra are labeled with the laser spot position and with the values of the radial and circumferential strain components. The solid lines follow the energy shift of the direct (*A*, black line) and indirect (*I*, red line) exciton transition.

To begin with, the micro-Raman measurements described in the Supplemental Material[16] show a progressive softening of the in-plane and out-of-plane vibrational modes while moving from the edge towards the center of the dome, in agreement with the expected tensile-strain increase[13,26,27]. However, the full extent of the effects of strain on the optoelectronic properties of TMD MLs can only be appreciated by looking at the dome's PL emission. Fig. 2(b) depicts a room temperature micro-PL scan taken along a diameter of the dome displayed in the inset of panel (a). The vertical axis indicates the energy of the emitted photons, while the base-10 logarithm of the PL intensity is shown in a false color scale. On moving from the edge toward the summit of the dome, the marked red-shift of the emission wavelength is accompanied by an equally striking decrease (about a factor of 20) of the PL intensity. Figure 2(c) describes in more detail the dramatic changes of the emission spectra from the dome's edge ($r = -2.85$ μm) to the dome's center ($r = -0.06$ μm). Each spectrum is labelled also with the pertinent values of the radial and circumferential strain components [see panel (a)]. The micro-PL spectra recorded close to the edge are dominated by the direct ($K_{CB}$-$K_{VB}$) band gap exciton (*A*), whose energy (equal to 2.00 eV in a strain-free reference $WS_2$ ML[8]) is red-shifted by the tensile strain exerted on the dome. As the excitation laser moves toward the center, the direct exciton keeps red-shifting and concomitantly a new, less intense band, labeled *I*, takes over and eventually dominates the spectrum. We ascribe this band to the $K_{CB}$-$\Gamma_{VB}$ *indirect* band gap exciton. In fact, as predicted by numerous theoretical works[5,18,19,21,22,23,24,25] the presence of strain in TMD MLs should result in a significant reordering of the energies of the critical points of the band structure. In particular, for tensile biaxial strains $\varepsilon \gtrsim 1\%$ in $WS_2$[18,19,21,22,23,25], the valence band maximum should change from the K to the $\Gamma$ point of the reciprocal space. Even though this change (*i*) is expected to occur for values of $\varepsilon$ that are well within reach of current strain modulation techniques[28,29,30,31], and (*ii*) should result in rather dramatic variations of the optical properties of the material, the currently available experimental evidence of this direct-to-indirect transition is either not particularly apparent[32,33] or absent[13,29,30,31]. This is possibly due to a less-than-perfect adhesion between the sample and the strain-inducing devices employed in some of the previous studies, resulting in an incomplete transfer of the applied stress to the TMD ML. In the present work, however, this is not an issue, as large biaxial strains—in the range between 1 - 3% —are induced by the pressure exerted on the TMD ML by the $H_2$ gas trapped and perfectly sealed within the dome.

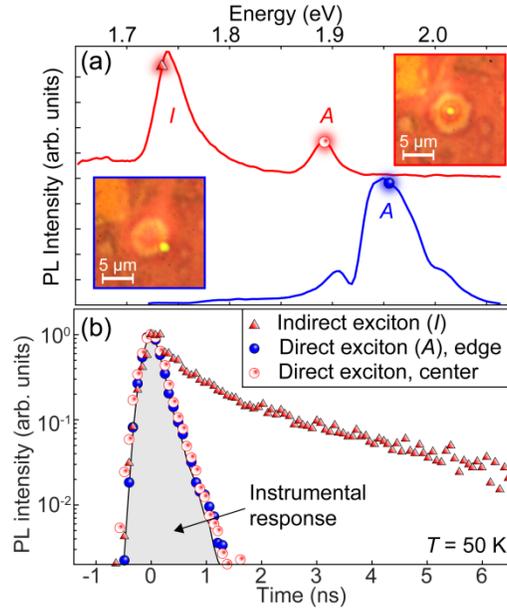

FIG.3. (a) $T$=50 K micro-PL spectra recorded at the center (red) and edge (blue) of a $WS_2$ dome. The insets are optical microscope images of the dome, showing the laser spot position corresponding to each spectrum. The symbols superimposed to the spectra indicate the energy at which the signal temporal decay shown in panel (b) was recorded. (b) Temporal evolution of the micro-PL signal relative to the specific photon energy and position on the dome highlighted in panel (a). The gray-shaded area refers to the exciting laser decay curve and sets the temporal resolution.

To confirm the previous attributions, we investigated the temporal decay of the micro-PL signal of $WS_2$ domes. These were cooled down at 50 K to minimize the contribution of non-radiative decay channels[34]. Interestingly, the reduction of the dome's volume at cryogenic temperatures[8]—due to the contraction of the $H_2$ gas trapped inside the dome—is nearly brought to a halt by the deposition of a thin methylpentane layer on the sample surface (see Supplemental Material), thus making it possible to spatially resolve the PL signal from different zones of the dome. Fig. 3(a) shows the micro-PL spectra of a $WS_2$ dome recorded at the edge (where the *A* exciton dominates) and center (where the *I* exciton can be observed along with the red-shifted *A* exciton recombination); see pictures in the insets. Fig. 3(b) shows the micro-PL decay curve relative to the different transitions displayed in (a). Most notably, the *A* and *I* excitons exhibit largely different temporal behaviors: The decay time of the *A* exciton is instrument-limited (<250 ps) consistent with other reports[34,35]. Instead, the *I* exciton shows a much longer temporal decay that can be fitted by a double exponential function with two decay times equal to (0.40 ± 0.06) ns and (2.9 ± 0.7) ns, which clearly points to an indirect optical transition[35].

We now establish the strain conditions that induce the K-Γ crossover in the VB. This is an especially important aspect with regard to the optoelectronic properties of TMD MLs and to the enormous potential that mechanical stress holds to engineer those properties. For instance, application of a seamless gradient of strain in these materials could be exploited as an efficient broad-band concentrator of photogenerated carriers in flexible solar cells[5]. Nevertheless, the occurrence of a strain-induced transition in the band gap character may affect both the absorption/emission properties and the carrier dynamics characteristics of devices based on TMD MLs. Furthermore, in the present case, the strain gradient enables the spatial concentration of long-lived *k*-indirect excitons with potential benefit for creating excitonic Bose condensates[36].

Figure 4(a) illustrates a micro-PL experiment performed on a single dome, highlighting the relevant physical processes discussed next. Panel (b) shows the peak energy $E_{A,I}$ of the *A* and *I* excitonic transitions derived from the same dome of Fig. 2 as a function of the "in-plane" strain $\varepsilon_p = \varepsilon_r + \varepsilon_t$. This choice is grounded on the hypothesis that each of the two planar strain components brings a similar effect on the ML band gap[18]. Moreover, for biaxially strained TMDs $\varepsilon_z = -\frac{D_{13}}{D_{33}} \varepsilon_p$, where $D_{13}$ and $D_{33}$ are the pertinent components of the elasticity matrix[16,37]. Thus, as discussed in Supplemental Material[16] the strain dependence of the energies of the *A* and *I* excitons can be written as

$$E_{A,I}(\varepsilon_p) = E_{A,I}(0) - \Delta_{A,I} \cdot \varepsilon_p, \tag{1}$$

where $\Delta_{A,I}$ is the shift rate with strain of the *A* (*I*) exciton. To correctly interpret the data shown in Fig. 4(b), however, we also have to consider that the continuous variation of the strain field on the dome surface [Fig. 2(a)]—and hence the progressive decrease of the band gap energy from the dome edge toward its center—leads excitons to drift toward the minimum energy available within their diffusion length before recombining [see panel (a)][5,6].

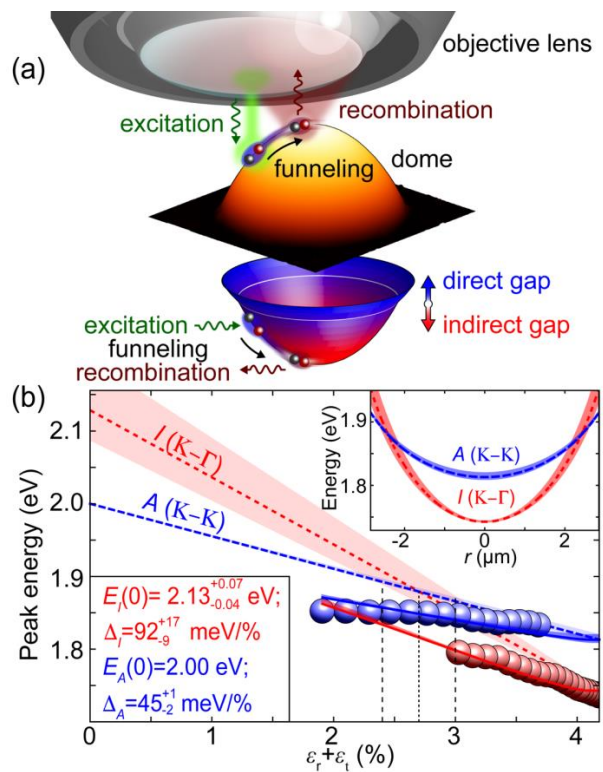

FIG. 4. (a) Sketch of a micro-PL experiment (excitation+recombination) on a single $WS_2$ dome (whose AFM image is shown in shaded orange), wherein an exciton drifts over the dome strain distribution (funnel effect[5,6]). The blue-red paraboloid provides a correspondence between the dome AFM image and the exciton energy, highlighting the direct-indirect transition region (see below). (b) Dependence of the energy of the direct (blue dots) and indirect (red dots) exciton transitions on the in-plane strain tensor, $\varepsilon_p = \varepsilon_r + \varepsilon_t$. The continuous blue/red lines (relative to the direct/indirect exciton) are fits based on Eq. (1), while also taking funneling [see panel (a)] into account. These fits entail linear dependences of the exciton energies on $\varepsilon_p$, which are displayed as dashed lines; the shaded areas enveloping each curve account for the uncertainty of our fitting procedure (see main text). In turn, these linear dependences yield the evolution of the direct and indirect exciton energy across the dome, plotted in the inset of panel (b) [the paraboloid sketched in (a) is also based on this evolution].

Such a funnel effect, combined with the finite exciting/collecting area of the objective, alters the correspondence between the coordinate $r$ (and thus $\varepsilon_p$) and the exciton energy derived from the emission spectra[6]. The solid curves displayed in Fig. 4(b) result from a fit performed by taking into account the exciton funneling [see Supplemental Material[16]], while fixing the radius of the collection area, $R_c$, to $2.5\sigma$ ($\sigma = 0.23$ μm is the laser spot size[12]). The actual [*i.e.*, free from the funnel effect; see Eq. (1)], "linearized" strain dependence of the *A* (*I*) exciton is shown in Fig. 4(b) as a blue (red) dashed line. The surrounding shaded areas—covering the regions spanned by the trends computed for $2\sigma \leq R_c \leq 3\sigma$—represent the uncertainty of our procedure. This analysis permits to set the direct-to-indirect band gap crossover point at $\varepsilon_p = (2.7\pm0.3)\%$ highlighted by vertical dashed lines in Fig. 4(b). Finally, the top-right inset in Fig. 4(b) provides the *A* and *I* exciton energy as a function of the dome radial coordinate. The displayed fits yield $\Delta_A = 45^{+1}_{-2}$ meV/$\varepsilon$%, $E_I(0) = 2.13^{+0.07}_{-0.04}$ eV and $\Delta_I = 92^{+17}_{-9}$ meV/$\varepsilon$% [$E_A(0)$ is fixed to 2.00 eV, the strain-free ML exciton energy]. These data compare rather favorably with the experimental[30,38], and theoretical[18,19,21,22,23,25,39] ones as discussed in Ref. 16. We finally point out that we observed similar findings also in $MoS_2$ and $WSe_2$[16]. This gives to our results a particular relevance regarding the general electronic properties of TMD 2D crystals.

In conclusion, we investigated the intertwined strain and electronic properties of spherically-deformed TMD monolayers. We observed that sufficiently high tensile in-plane strains ($\varepsilon_p \sim 2.7\%$ in $WS_2$-ML) turn a direct band gap material into an indirect-gap one. This general behavior, common to other TMDs—like $MoS_2$ and $WSe_2$—must be considered when 2D crystal are to be employed in flexible optoelectronic devices, or possibly exploited for the observation of quantum many-body effects involving long-lived *k*-space indirect excitons[36].

AUTHOR INFORMATION

[*] corresponding author: antonio.polimeni@roma1.infn.it

*Acknowledgements.* We thank A. Surrente for comments and suggestions. We acknowledge support by Sapienza Università di Roma under the grants "Ricerche Ateneo" years 2017 and 2018 (A.P. and M.F.). M.F. and G.P. acknowledge support and funding from the Italian Ministry for Education, University and Research within the Futuro in Ricerca (FIRB) program (project DeLIGHTeD, Prot. RBFR12RS1W). This project has also received funding from the European Union's Horizon 2020 research and innovation program No. 641899. We acknowledge fund support from Australian Research Council (ARC) (numbers DE140100805 and DP180103238), and ARC Centre of Excellence in Future Low-Energy Electronics Technologies (project number CE170100039).

[26]. F. Wang, I. A. Kinloch, D. Wolverson, R. Tenne, A. Zak, E. O'Connell, U. Bangert, and R. J. Young, Strain-induced phonon shifts in tungsten disulphide nanoplatelets and nanotubes, 2D Materials **4**, 015007 (2017).

[27]. A. M. Dadgar, D. Scullion, K. Kang, D. Esposito, E.-H. Yang, I. P. Herman, M. A. Pimenta, E.-J. G. Santos, and A. N. Pasupathy, Achieving large, tunable strain in monolayer transition-metal dichalcogenides, Chem. Mater. **30**, 5148 (2018).

[28] Q. Zhang,, Z. Chang, G. Xu, Z. Wang, Y. Zhang, Z.-Q. Xu, S. Chen, Q. Bao, J. Z. Liu, Y.-W. Mai, W. Duan, M. S. Fuhrer, and C. Zheng, Strain Relaxation of Monolayer $WS_2$ on Plastic Substrate, Adv. Funct. Mater. **26**, 8707-8714 (2016).

[29] R. Yang, J. Lee, S. Ghosh, H. Tang, R. M. Sankaran, C. A. Zorman, and P. X.-L. Feng, Tuning optical signatures of single- and few-layer $MoS_2$ by blown-bubble bulge straining up to fracture, Nano Lett. **17**, 4568 (2017).

[30]. R. Frisenda, M. Drüppel, R. Schmidt, S. M. de Vasconcellos, D. P. de Lara, R. Bratschitsch, M. Rohlfing, and A. Castellanos-Gomez, Biaxial strain tuning of the optical properties of single-layer transition metal dichalcogenides. 2D Materials **4**, 10 (2017).

[31]. H. J. Conley, B. Wang, J. I. Ziegler, R. F. Haglund Jr., S. T. Pantelides, and K. I. Bolotin, Bandgap engineering of strained monolayer and bilayer $MoS_2$. Nano Lett. **13**, 3626 (2013).

[32] J. Chaste, A. Missaoui, S. Huang, H. Henck, Z. Ben Aziza, L. Ferlazzo, C. Naylor, A. Balan, A. T. C. Johnson Jr., R. Braive, and A. Ouerghi, Intrinsic Properties of Suspended $MoS_2$ on $SiO_2$/Si Pillar Arrays for Nanomechanics and Optics, ACS Nano **12**, 3235-3242 (2018).

[33]. Y. Wang, C. Cong, W. Yang, J. Shang, N. Peimyoo, Y. Chen, J. Kang, J. Wang, W. Huang, and T. Yu, Strain-induced direct-indirect bandgap transition and phonon modulation in monolayer $WS_2$. Nano Res. **8**, 2562 (2015).

Supplemental Material for

# Variation of the fundamental band gap nature in curved two-dimensional WS$_2$

E. Blundo[‡], M. Felici[‡], T. Yildirim[†], G. Pettinari[◊], D. Tedeschi[‡], A. Miriametro[‡], B. Liu[†], W. Ma[†], Y. Lu,[†,#] and A. Polimeni[‡,*]

[‡] Dipartimento di Fisica, Sapienza Università di Roma, 00185 Roma, Italy

[†] Research School of Electrical, Energy and Materials Engineering, College of Engineering and Computer Science, The Australian National University, Canberra, ACT 2601, Australia

[◊] Institute for Photonics and Nanotechnologies, National Research Council, 00156 Roma, Italy

[#] ARC Centre of Excellence in Future Low-Energy Electronics Technologies (FLEET) ANU node, Canberra, ACT 2601, Australia

[*] antonio.polimeni@roma1.infn.it

**Abstract**

In this Supplemental Material we (*i*) provide details about the finite-element method employed to calculate the strain tensor over spherically-deformed WS$_2$ membranes or domes, (*ii*) evaluate the biaxial strain acting on the domes' summit according to a continuum mechanical model, (*iii*) show micro-Raman measurements over the dome surface to account for the strain gradient, (*iv*) report the effect of methylpentane capping on the temperature dependence of the dome volume, (*v*) explain how the funneling effect was taken into account to determine the Γ-K cross-over in the valence band, (*vi*) show results about the direct-to-indirect band gap transition in MoS$_2$ and WSe$_2$.

**Contents**

# 1. FEM profile and strain modelling

The height profile of the domes and the evolution of the strain tensor across the domes' surface were modelled via finite-element method (FEM) calculations in the framework of nonlinear-membrane theory were employed.[1] The presence of hydrogen gas within the domes following proton irradiation is modelled as an internal, upward pressure load acting on the flat membrane surface. To run the calculations, the radius of the dome, the pressure of the hydrogen gas and the elasticity matrix (as accounted for below) of the compound are given as input parameters to simulate the interplay between adhesion forces, gas expansion and elastic forces. The pressure value is chosen so that the simulated dome maximum height matches that one measured by AFM. The relatively small size of the domes allows us to use membrane theory and to neglect the effects of bending stiffness. Moreover, due to the relatively large dome heights compared to the thickness of a $MX_2$ monolayer (ML), we employ a nonlinear Green-Lagrange strain tensor. Thanks to the mirrored symmetry about the longitudinal axis of the domes, an axisymmetric formulation using polar coordinates is used. Thus, a model implemented in COMSOL Multiphysics 5.1 is employed for the simulations of the dome formation. A line element is used to simulate the membrane with the starting thickness of a $MX_2$ ML. One end of the line element is subjected to a fixed constraint (*i.e.* null displacement) while the rest of the line is free to move in the longitudinal direction. An extra fine mesh is used for all simulations and a 0.001 convergence setting is used to ensure the numerical accuracy of the solutions. A constant Newtonian solver is used for the numerical solver procedure. Due to the anisotropic stiffness of $MX_2$ compounds, an anisotropic stiffness matrix was implemented in the linear elastic material node[2,3,4].

The following anisotropic elasticity matrices[2,3] (all units in GPa) were here employed.

For MoS$_2$,

$$D = \begin{pmatrix} 178 & 45 & 1 & 0 & 0 & 0 \\ 45 & 178 & 1 & 0 & 0 & 0 \\ 1 & 1 & 3 & 0 & 0 & 0 \\ 0 & 0 & 0 & 1 & 0 & 0 \\ 0 & 0 & 0 & 0 & 1 & 0 \\ 0 & 0 & 0 & 0 & 0 & 67 \end{pmatrix}$$

For WSe$_2$,

$$D = \begin{pmatrix} 158 & 31 & 1 & 0 & 0 & 0 \\ 31 & 158 & 1 & 0 & 0 & 0 \\ 1 & 1 & 2 & 0 & 0 & 0 \\ 0 & 0 & 0 & 1 & 0 & 0 \\ 0 & 0 & 0 & 0 & 1 & 0 \\ 0 & 0 & 0 & 0 & 0 & 64 \end{pmatrix}$$
.

In the case of WS$_2$, only the matrix elements $D_{12}$, $D_{22}$, and $D_{66}$ could be found in Refs. 2, 3 and 5. However, the other elements are almost the same for all the compounds. Due to the analogous effect induced by strain on the lattice constant[6], for the unknown elements the same elements of the MoS$_2$ matrix were used, so that the following anisotropic elasticity matrix was used for the WS$_2$ simulations.

$$D = \begin{pmatrix} 196 & 43 & 1 & 0 & 0 & 0 \\ 43 & 196 & 1 & 0 & 0 & 0 \\ 1 & 1 & 3 & 0 & 0 & 0 \\ 0 & 0 & 0 & 1 & 0 & 0 \\ 0 & 0 & 0 & 0 & 1 & 0 \\ 0 & 0 & 0 & 0 & 0 & 76 \end{pmatrix}$$

We point out that the Poisson's ratios calculated from the values reported in these matrices ($\nu = D_{12}/D_{22}$) are in excellent agreement with the values reported in other works (see Supplementary Note 2).

The mechanical model here described allows us to reproduce the dome shape and calculate the principal components of the strain tensor, as displayed in Fig. S1 for $WS_2$, $MoS_2$, and $WSe_2$. The use of the principal components ensures that the strain tensor is a diagonal matrix, i.e., there are no shear-strain components to take into account. A sketch of the three principal components is given in Fig. 2a of the main text. The tensor is characterized by two in-plane components: The circumferential ($\varepsilon_t$) and radial ($\varepsilon_r$) components -analogously defined to Ref. 1– and by an out-of-plane component: The perpendicular strain component ($\varepsilon_z$). Moreover, for biaxially strained TMDs, the perpendicular component of the strain tensor can be written as[7] $\varepsilon_z = -\frac{D_{13}}{D_{33}} \cdot (\varepsilon_r + \varepsilon_t)$, where $D_{13}$ and $D_{33}$ are the pertinent elements of the elasticity matrix reported above. Notice that this implies that $\varepsilon_z < 0$, accounting for a compression in the out-of-plane direction that increases as ($\varepsilon_r + \varepsilon_t$) increases.

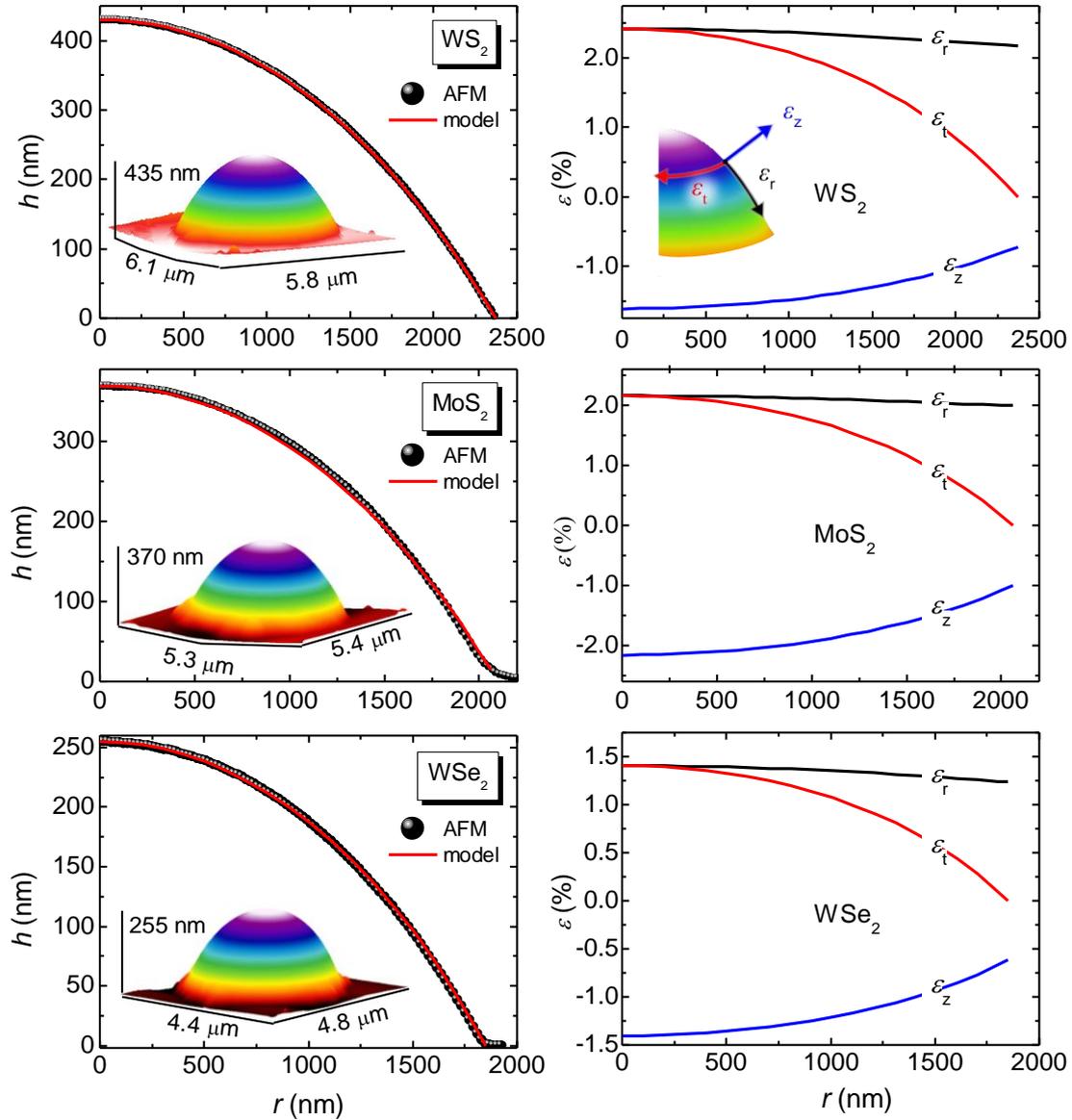

**Figure S1. Mechanical model vs experiment.**
**Left panels:** Experimental dependence of the dome height $h$ on the radial coordinate $r$ (points) for domes formed by different TMD materials. The $h$-$r$ plots were derived from the atomic force microscope images shown as insets. The red lines are the results of the mechanical model employed to reproduce the dome shape and calculate the strain tensor components displayed on the right panels. **Right panels:** Dependence of the principal components of the strain tensor on the radial coordinate $r$ of the domes. The definition of the radial ($\varepsilon_r$) and circumferential ($\varepsilon_t$) in-plane components and of the perpendicular ($\varepsilon_z$) out-of-plane strain component are depicted in the inset of the first panel on the right.

The results obtained by means of this mechanical model for the strain tensor are in good agreement with Hencky's model (described in the next paragraph). In Table 1 we report a comparison between Hencky's model and our model, for the 3 domes of Fig. S1.

|  | Hencky's model[8,9] | this work |
|---|---|---|
| $WS_2$ | 2.40% | 2.43% |
| $MoS_2$ | 2.13% | 2.16% |
| $WSe_2$ | 1.40% | 1.40% |

**Table 1:** Comparison between the expected values of the strain at the dome summit evaluated by Hencky's model[8,9] $\varepsilon_m = \varepsilon_t = \varepsilon_r = f(v) \cdot (h_m/R)^2$ (see Supplementary Note 2) and those derived by our mechanical model. The values reported in the table refer to the domes whose data are shown in Figure S1.

Besides the height and strain profile, the mechanical model here described also allows us to estimate the internal pressure of the gas contained within the domes. The estimation of the pressure dependence on the dome footprint radius is displayed in Fig. S2 for $WS_2$ domes.

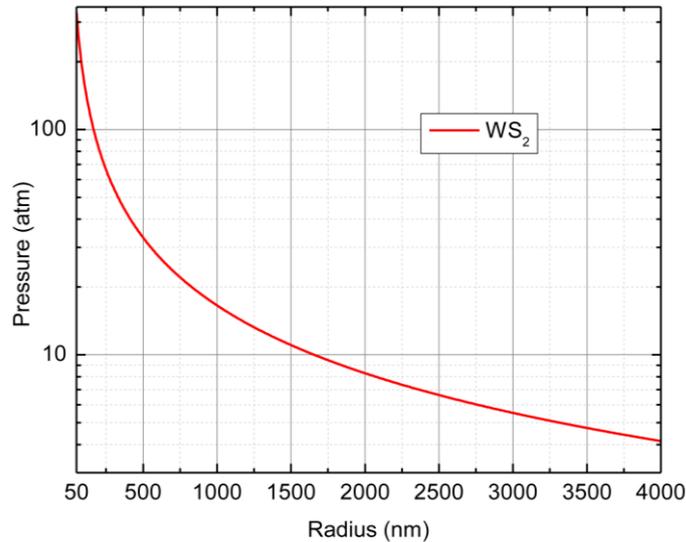

**Figure S2. Internal pressure estimation.**
Predicted dependence of the $H_2$ pressure as a function of the dome footprint radius in $WS_2$ domes. The data are evaluated by using the same mechanical model employed for simulating the dome profile and determining the strain tensor components.

## 2. Evaluation of biaxial strain at the dome summit (Hencky's model)

The expression for biaxial strain at the top of the dome is accounted for in Refs. 8 and 9 and is given by the Hencky's formula:

$$\varepsilon_m = \left(\frac{h_m}{R}\right)^2 f(v) = \left(\frac{h_m}{R}\right)^2 \frac{b_0(v)(1-v)K(v)^{2/3}}{4},$$

where $h_m$ and $R$ are the maximum height and the footprint radius of the dome, respectively. $b_0$ and $K$ are two quantities that depend only on the Poisson's ratio $v$. The values of $v$ for all the materials illustrated in Fig. S1 are listed in the following table[10].

|   | $WS_2$ | $MoS_2$ | $WSe_2$ |
|---|---|---|---|
| $v$ | 0.217 | 0.249 | 0.192 |

**Table 1. Poisson's ratios.**
Poisson's ratios for the six compounds (after Ref. 10).

$b_0$ and $K$ can be evaluated by numerical methods. Therefore, to estimate them we perform an interpolation between the values reported in the literature[9,11] from which we finally get $f(v)$ for all the compounds shown in Fig. S1. The values of $b_0$, $K$, and $f(v)$ employed in this work are listed in the following table.

|   | $WS_2$ | $MoS_2$ | $WSe_2$ |
|---|---|---|---|
| $b_0$ | 1.689 | 1.702 | 1.680 |
| $K$ | 3.274 | 3.386 | 3.192 |
| $f(v)$ | 0.729 | 0.721 | 0.736 |

**Table 2. $b_0$, $K$, and $f(v)$.**
Values of $b_0$, $K$, and $f(v)$ calculated via an interpolation between the values reported in the literature[9,11].

Finally, the average values of the $h_m/R$ ratio and of $\varepsilon_m$—as obtained from the analysis of the data presented in Fig. S1—are reported in the following table for all the compounds investigated here:

|       | WS$_2$       | MoS$_2$      | WSe$_2$      |
|-------|--------------|--------------|--------------|
| $h_m/R$ | 0.16±0.02   | 0.16±0.02    | 0.15±0.01    |
| $\varepsilon_m$ | (2.0±0.4)% | (1.8±0.4)% | (1.6±0.2)% |

**Table 3. Aspect ratios and strain at the summit.**

First row: Average values of the aspect ratios ($h_m/R$) for the three compounds.

Second row: Average values for the strain at the summit, calculated for the three compounds according to Hencky's formula and using the aspect ratios in the first row and the values of $f(v)$ reported in Table 2.

## 3. Micro-Raman measurements

The following figure shows a micro-Raman mapping performed along the diameter of the same WS$_2$ dome, whose micro-PL mapping is shown and discussed in Fig. 2 of the main text. Details about micro-Raman measurements are reported in the caption of following Figure S3.

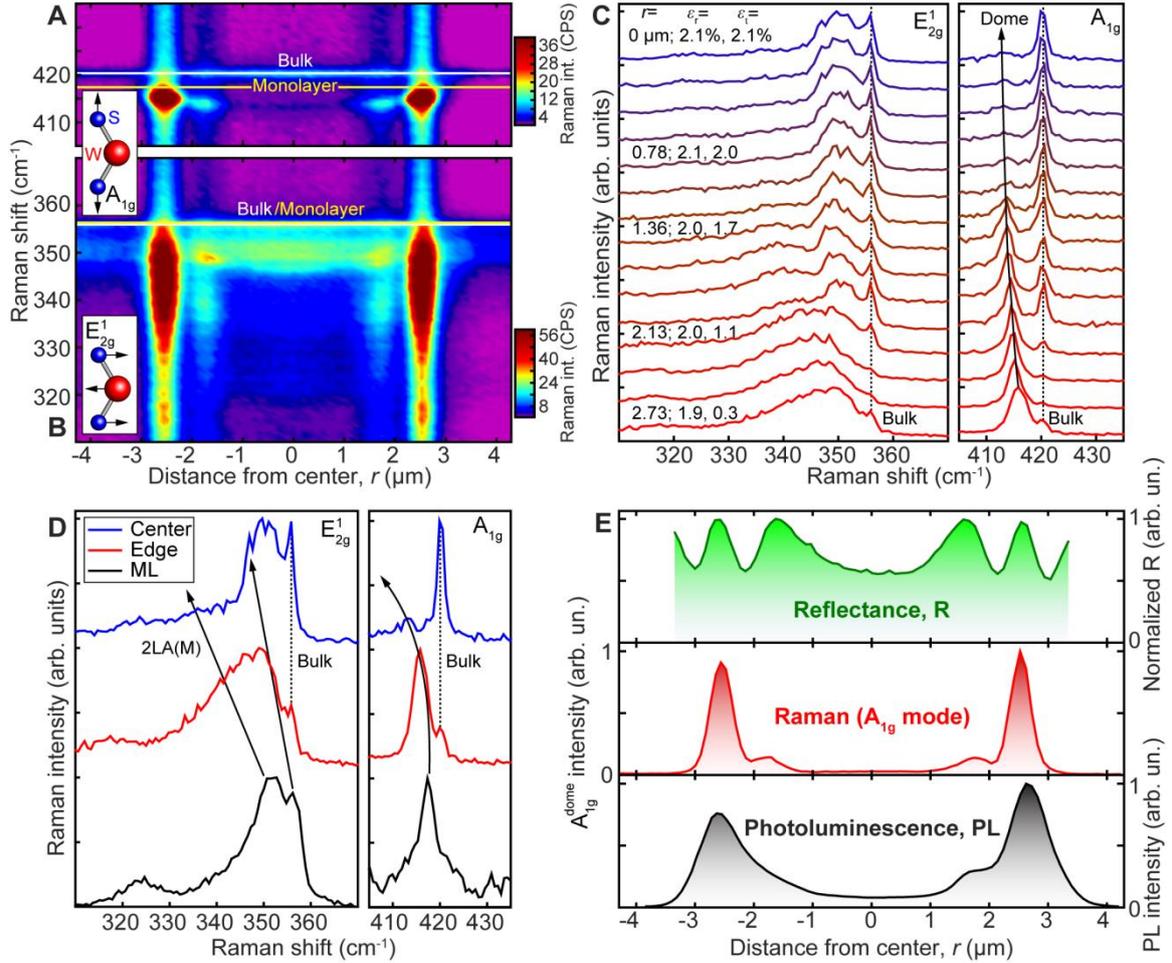

**Figure S3. μ-Raman mapping of a WS$_2$ dome**

(**A**) One-dimensional, room-temperature ($T = 297$ K) micro-Raman scan across a WS$_2$ dome, in the spectral region of the A$_{1g}$ phonon mode (the dome is the same one displayed in Fig. 2a of the main text). The horizontal axis indicates the laser spot position with respect to the dome center ($r$), whereas the vertical axis indicates the Raman shift with respect to the laser line (laser wavelength $\lambda$ = 532.2 nm). The micro-Raman intensity is shown in a false color scale (see colorbar). The white (yellow) line marks the position of the A$_{1g}$ mode in bulk (monolayer) WS$_2$[12,13]. **Inset:** Atomic displacements associated with the A$_{1g}$ mode (the vertical axis of the sketch is perpendicular to the monolayer plane). (**B**) Same as Panel A, but for the E$^1_{2g}$ mode. (**C**) **Left:** Evolution of the normalized micro-Raman spectrum of the dome ($T = 297$ K) in the spectral region of the E$^1_{2g}$ mode, as the laser spot is scanned from the dome's right edge (bottom) to its apex (top). Some selected spectra are labeled with the position of the laser spot and with the values of the radial ($\varepsilon_r$) and circumferential ($\varepsilon_t$) components of the strain tensor (see Fig. 2a of the main text and Supplementary Fig. S1). The dotted line marks the position of the $E^1_{2g}$ mode in bulk WS$_2$. **Right:** Same as on the left, but for the $A_{1g}$ mode. The black arrow follows the mode shift as the laser is scanned across the dome. **(D) Right:** Comparison between the normalized micro-Raman spectrum ($T = 297$ K) in monolayer WS$_2$ (bottom) and at the edge (middle) and in the center (top) of the dome, in the spectral region of the $E^1_{2g}$ mode. The dotted line marks the position of the $E^1_{2g}$ mode in bulk WS$_2$; the arrows follow the Raman shifts of the $E^1_{2g}$ and 2LA(M) modes. The mode at 330 cm$^{-1}$ in the ML spectrum was also observed in Ref. 14. We tentatively ascribe this mode to a LA replica. Right: Same as on the left, but for the $A_{1g}$ mode. **(E) Bottom:** Dependence of the integrated micro-PL intensity on the position of the laser spot. In order to obtain the displayed intensity profile, the micro-PL spectra displayed in panels b-c of Fig. 3a in the main text were integrated between 1.7 and 1.95 eV. **Middle:** Evolution of the intensity of the $A_{1g}$ Raman mode as the laser is scanned across the dome. The reported intensity values were obtained by fitting each Raman spectrum (see panel A and the right-hand side of panel C) with the function $I_{tot}= I_{dome}+I_{bulk}+I_{bkg}$. Here, $I_{tot}$ is the total spectrum, whereas $I_{bkg}$ is a flat background. $I_{dome}$ and $I_{bulk}$ are Gaussian functions, respectively taking into account the Raman peaks associated with lattice vibrations in the dome layer and in the underlying bulk WS$_2$. In the panel we report $I_{dome}+I_{bkg}$, thereby excluding the contribution of bulk WS$_2$ to the spectrum. **Top:** Normalized reflectance at 532.2 nm as a function of the position of the laser spot. The displayed profile was obtained by collecting the light reflected by the sample as the laser was scanned across the dome, under the same experimental conditions used for micro-Raman and micro-PL measurements (laser wavelength $\lambda = 532.2$ nm, $T = 300$ K, confocal configuration). Note the much different dynamic range spanned by the three intensity profiles with a ratio between the maximal and minimal value equal to 13 and 33 for the micro-PL and micro-Raman profiles, respectively. For micro-PL it is straightforward to ascribe this intensity reduction to the direct-to-indirect band gap transition taking place as one moves from the edge to the center of the dome (see main text, Fig. 2). An analogous behavior to that of the dome discussed in Fig. 2b,c is found also for other WS$_2$ domes despite of their different dimensions, as also attested by the same ring-like pattern of the laser-excited red luminescence observed for all the domes in Fig. 1c, which suggests a minor role to be played by interference. It seems reasonable to assume a similar origin for the observed reduction of the micro-Raman signal at the dome's center. In this case, some interferential effects can be noticed in between the edge and the center of the dome, where an analogous modulation to that of the reflectance profile can be seen (see, in particular, the minimum for $r \sim \pm 2$ µm). We can therefore conclude that the modulation of the micro-PL and micro-Raman profiles are chiefly due to the strain-induced variations of the electronic properties across the surface of our domes, and interference does not play a primary role.

**4. Effects of methylpentane deposition on the *T* dependence of the dome's size**

As reported in Ref. 15, the contraction of the $H_2$ gas trapped inside a dome leads to a progressive reduction of the dome's volume at cryogenic temperatures, which culminates in the dome's disappearance upon reaching the vapor-to-liquid transition temperature at about 32 K. Even though this phenomenon is fully reversible (each dome reappears in its original position when the temperature is increased), the dome's shrinkage at cryogenic temperatures makes it increasingly difficult to spatially resolve the micro-PL signal from different zones of the dome. As noted in the main text, this is potentially highly problematic for time-resolved micro-PL measurements, which must be performed at low temperature to fully appreciate the existing differences in the temporal behavior of the direct and indirect exciton. At room temperature, indeed, non-radiative decay channels dominate the exciton dynamics, as also reported, *e.g.*, in Ref. 16. As illustrated in Fig. S4, however, this issue can be conveniently overcome by covering the sample surface with a thin layer of methylpentane. Indeed, the adhesion of the latter to the dome's walls is enough to nearly stop the dome's contraction with decreasing *T*, without sizably altering the dome's emission properties (see main text).

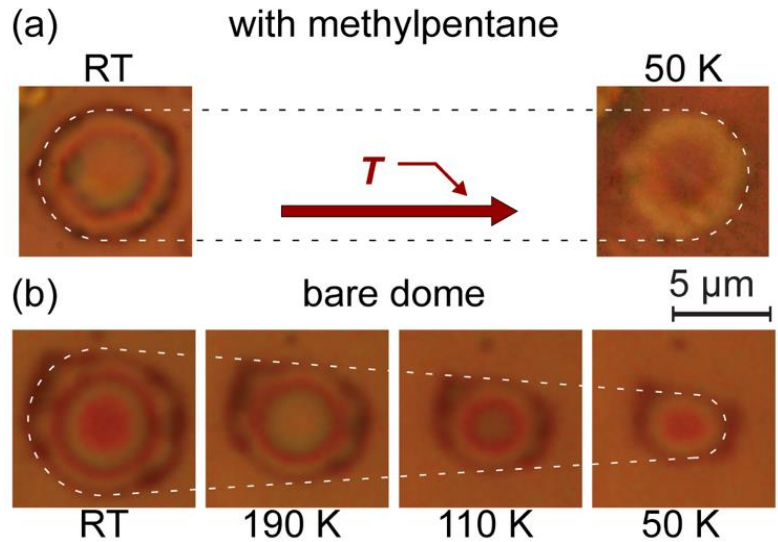

**Figure S4. Slowing down the dome's contraction at low *T* by methylpentane deposition**

(a) Optical microscope images of the WS$_2$ dome on which the time-resolved micro-PL measurements discussed in the main text (see Fig. 3) were acquired. The image on the left was acquired at room temperature (*T*~290 K), whereas the image on the right was taken at *T*=50 K (*i.e.,* the temperature at which the time-resolved micro-PL measurements were performed). Even though the same 100× objective (NA=0.75) was used for both images, the quality of the one acquired at *T*=50 K is affected by the aberrations due to the presence of the optical cryostat window in between the objective and the sample. Nevertheless, the effects of methylpentane deposition are clearly visible, *i.e.*, the dome's size becomes nearly insensitive to temperature changes. This is in sharp contrast with the situation depicted in panel (b), which displays the *T* dependence of the dome's size in a "bare" (*i.e.*, not covered with methylpentane) sample. The dome's shrinkage with decreasing *T* is clearly visible in the displayed optical images (acquired with a 50× objective with NA=0.5).

**5. Direct-to-indirect band gap transition and funneling effect**

First of all, we justify Eq. (1) of the main text, according to which

$$E_{A,I}(\varepsilon_p) = E_{A,I}(0) - \Delta_{A,I} \cdot \varepsilon_p. \tag{S1}$$

In particular, we need to clarify why the energies of the *A* and *I* excitons only depend on the total in-plane strain, $\varepsilon_p = \varepsilon_r + \varepsilon_t$, rather than being an explicit function of the three principal components of the strain tensor, namely

$$E_{A,I}(\varepsilon_r, \varepsilon_t, \varepsilon_z) = E_{A,I}(0) - \left(\Delta_{A,I}^r \cdot \varepsilon_r + \Delta_{A,I}^t \cdot \varepsilon_t + \Delta_{A,I}^z \cdot \varepsilon_z\right) \tag{S2}$$

(it should be noted that the use of the principal components implies that the strain tensor is a diagonal matrix, *i.e.*, there are no shear-strain components to be taken into account). According to Ref. 6, the energy shifts of the first Brillouin-zone critical points of TMD MLs induced by the presence of an in-plane uniaxial tensile strain do not depend heavily on the direction along which the deformation is exerted, so that we can write $\Delta_{A,I}^r \cdot \varepsilon_r + \Delta_{A,I}^t \cdot \varepsilon_t \sim \Delta_{A,I}^p \cdot (\varepsilon_r + \varepsilon_t) = \Delta_{A,I}^p \cdot \varepsilon_p$. Moreover, as we noted in the main text, for biaxially strained TMDs the perpendicular component of the strain tensor can be written as $\varepsilon_z = -\frac{D_{13}}{D_{33}} \cdot \varepsilon_p$, where $D_{13}$ and $D_{33}$ are the pertinent components of the elasticity matrix reported in Section 1 of this supplemental Material[17]. As a consequence, Eq. (S2) can be rewritten as

$$E_{A,I}(\varepsilon_p) = E_{A,I}(0) - \left(\Delta_{A,I}^p \cdot \varepsilon_p - \Delta_{A,I}^z \cdot \frac{D_{13}}{D_{33}} \varepsilon_p\right), \tag{S3}$$

which is identical to Eq. (S1) (*i.e.*, to Eq. (1) of the main text), with $\Delta_{A,I} = \Delta_{A,I}^p - \Delta_{A,I}^z \cdot \frac{D_{13}}{D_{33}}$. The validity of Eq. (1) of the main text implies that the analysis of the dependence of $E_{A,I}$ on $\varepsilon_p$ can yield important information on the energy bands of WS$_2$—such as the energy of the *I* exciton for zero strain, $E_I(0)$—and on their dependence on strain, quantified through the $\Delta_{A,I}$ energy shift rates. In order to extract this information from the experimental data reported in Fig. 4 of the main text, however, we have to take into account the continuous reduction of the energy gap of the material on going from the edge to the center of the curved dome's surface. As a consequence of this reduction, the photogenerated carriers drift towards the minimum energy available within their diffusion length (funnel effect)[18,19], *i.e.*, towards the dome's center. Combined with the finite exciting/collecting area of the objective, this alters profoundly the correspondence between the coordinate $r$ (and thus $\varepsilon_p$) and the exciton energy resulting from the emission spectra. The curves superimposed on the experimental data in Fig. 4 simulate the local values of the energies of the *A* and *I* exciton band gaps. These latter were deduced by a model in which exciton annihilation takes place in the minimum energy available within the objective collection area[19]. For each set of data, the solid line results from a fit performed by fixing the radius of the collection area, $R_c$, to $2.5\sigma$ ($\sigma = 0.23$ μm is the laser spot size; see Ref. 12 in the main text), whereas the shaded areas are delimited by the fitting curves (barely observable) obtained for $R_c = 2\sigma$ and $3\sigma$. This range of $R_c$—which ultimately sets the uncertainty of our fitting procedure—was chosen to account for the effect of the pinhole used to spatially filter the collected light in our confocal microscope, which was indeed verified to transmit between 95% and 99% of the laser light reflected by the sample. The actual (*i.e.*, free from the funnel effect) strain dependences of the *A* and *I* excitons are provided in Fig. 4b by blue and red dashed lines—again computed for $R_c = 2.5\sigma$—and by their corresponding shaded areas, indicating the regions spanned by the theoretical trends in the $2\sigma \leq R_c \leq 3\sigma$ range.

|  | Our work | Literature |
|---|---|---|
| $\Delta_A$ (meV/%) | 45 | 46[20], 47[21] |
| $\Delta_I$ (meV/%) | 2.0 $\Delta_A$ | 1.1·$\Delta_A$[22], 2.1·$\Delta_A$[23] |
| $E_I(0) - E_A(0)$ (meV) | 130 | 117[22], 125[23], 173[24] |
| $\varepsilon_{transition}$ (%) | 2.7 | ≥2[6,25,26] |

**Table 1. comparison with literature.**
Shift rate values $\Delta$ of the direct (A) and indirect (I) exciton energy. $E_{A,I}(0)$ indicates the direct and indirect exciton energies at zero strain. In order to make meaningful a comparison with theoretical papers (where the absolute value of the band gap is underestimated), we report the difference between these quantities.

## 6. Direct-to-indirect band gap transition in MoS$_2$ and WSe$_2$ monolayers

The findings reported for WS$_2$ are general and were observed also in other TMD compounds. In the following figure S5, we show the micro-PL spectra recorded in different points of a single MoS$_2$ (left) and WSe$_2$ (right) dome, showing the dramatic changes of the emission spectrum on going from the dome's edge to its center.

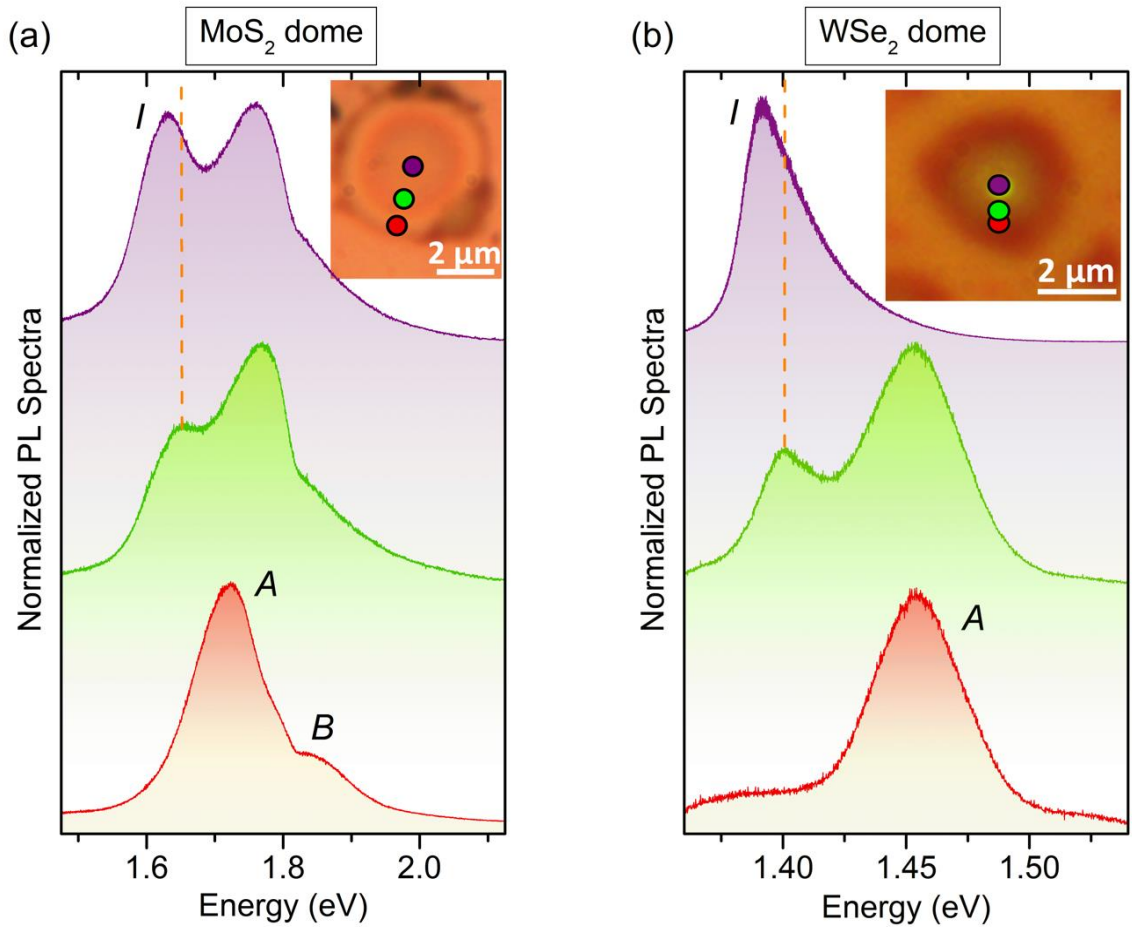

**Figure S5. Direct-to-indirect bandgap crossover in MoS$_2$ and WSe$_2$ domes.**
**(a)** micro-PL spectra acquired in the positions highlighted by the colored dots in the MoS$_2$ dome (with footprint radius $R = (2.46\pm0.06)$ µm) shown as inset. The spectra are peak-normalized for ease of comparison. At the edges of the dome (red dot in the inset) the direct $A$ exciton (corresponding to the K$_C$-K$_V$ transition) dominates the spectrum (red spectrum) and the $B$ exciton involving the conduction band minimum at K$_C$ and the lower split band at K$_V$ in valence band can also be observed. According to our mechanical model, at the edges of MoS$_2$ domes the strain is strongly anisotropic with a radial component equal to ~ 2 % and an almost null circumferential component (see Supplementary Figure S1). The $A$ peak appears to be broad and composite due to the complex strain distribution coupled to the finite resolution of our optical system (as discussed in the main text) and to funneling phenomena. These factors are also likely responsible for irregular behaviors close to the edges, such as the unexpected blueshift (of ~ 30 meV) observed while moving towards the dome center (green dot in the inset and corresponding green spectrum). Notice that deviations from the expected behavior towards higher energies were similarly observed also for uniaxially-strained $MoS_2$ monolayers in Ref. 27, for strains $\gtrsim 1.4$ %. The *A* band then redshifts while further going towards the center, where strain features a more regular biaxial distribution with $\varepsilon_r \approx \varepsilon_t$. In between the edges and the center (see green dot in the inset and corresponding green spectrum) a new band (*I*) appears, which we attribute to the indirect $K_C$-$\Gamma_V$ transition. This band starts dominating the spectrum at the dome summit (purple dot in the inset and corresponding purple spectrum). Notice that at the center the *I* and *A* band are redshifted by about 20 meV and 10 meV, respectively, with respect to the green dot, accordingly to the increase of biaxial strain and the theoretically predicted higher shift rate for the $K_C$-$\Gamma_V$ transition with respect to the $K_C$-$K_V$ transition[22,23]. The dashed orange line highlights the redshift of the *I* exciton. Concomitantly, a reduction in the PL signal is observed, the PL signal at the edge being ~ 10 times more intense than at the center. Based on our µ-PL measurements on several $MoS_2$ domes, the bandgap crossover is found to occur for this compound for an in-plane strain ~ 3-4 %. **(b)** Same as in panel (a) for the $WSe_2$ dome (with footprint radius $R = (1.42\pm0.06)$ µm) shown as inset. The direct *A* exciton (corresponding to the $K_C$-$K_V$ transition) dominates the spectrum at the edges (red dot in the inset and corresponding red spectrum). The indirect *I* band appears while moving towards the center (green dot and corresponding green spectrum), redshifts while approaching the dome summit (as highlighted by the dashed orange line) and finally dominates the spectrum (purple dot and spectrum). For this compound, no significant quenching of the PL signal is observed for increasing tensile strain, the signal often increasing while going from the edges towards the domes' summit. An increase in the PL emission was also observed in Ref. 28 for $WSe_2$ monolayer under uniaxial strain. In our domes, however, the signal is observed not to decrease even in presence of the direct-to-indirect bandgap transition (for the dome here shown the signal at the center is found to be almost 10 times higher than at the edges). This enhancement in the PL efficiency is likely aided by funneling effects combined with the small dimensions of the $WSe_2$ domes we can create: The larger domes have dimensions comparable to that of the excitation laser spot, resulting in funneling of excitons at the dome summit and thus favoring the indirect transition. For $WSe_2$, the bandgap crossover is found to occur for an in-plane strain ~ 2-3 %.